# Voltage Scaling of Graphene Device on SrTiO$_3$ Epitaxial Thin Film

*Jeongmin Park,[†,‡,⊥] Haeyong Kang,[‡,⊥] Kyeong Tae Kang,[†,§,⊥] Yoojoo Yun,[†,‡] Young Hee Lee,[†,‡,§] Woo Seok Choi,[§,*] Dongseok Suh[†,‡,*]*

[†] IBS Center for Integrated Nanostructure Physics, Institute for Basic Science, [‡] Department of Energy Science, [§] Department of Physics, Sungkyunkwan University, Suwon 440-746, Korea

[⊥] These authors contribute equally to this work.
[*] Corresponding authors: choiws@skku.edu; energy.suh@skku.edu

ABSTRACT

Electrical transport in monolayer graphene on $SrTiO_3$ (STO) thin film is examined in order to promote gate-voltage scaling using a high-$k$ dielectric material. The atomically flat surface of thin STO layer epitaxially grown on Nb-doped STO single-crystal substrate offers good adhesion between the high-$k$ film and graphene, resulting in nonhysteretic conductance as a function of gate voltage at all temperatures down to 2 K. The two-terminal conductance quantization under magnetic fields corresponding to quantum Hall states survives up to 200 K at a magnetic field of 14 T. In addition, the substantial shift of charge neutrality point in graphene seems to correlate with the temperature-dependent dielectric constant of the STO thin film, and its effective dielectric properties could be deduced from the universality of quantum phenomena in graphene. Our experimental data prove that the operating voltage reduction can be successfully realized due to the underlying high-$k$ STO thin film, without any noticeable degradation of graphene device performance.

Graphene has become one of the standard reference materials for the research of two-dimensional structured materials and devices over the past decade.[1-4] As several intriguing characteristics of graphene have been discovered,[5-7] the combination of graphene with other functional materials is now attracting much attention.[8-15] In particular, transition metal oxides with a high dielectric constant (high-$k$) or ferroelectric properties have been considered as promising candidates to study synergetic effects, such as ultrahigh doping, when integrated with two-dimensional materials.[13-22] However, these anticipated operational characteristics of designed devices are difficult to realize because charge conduction in graphene is very sensitive to its environment due to the two-dimensional nature itself, as well as its nonintimate contact with the oxide's surface.[23] Specifically, surface defects and/or adsorbates have been considered as the sources of such limitations.[24,25]

To study charge transport through graphene under the influence of functional oxides, graphene devices have been fabricated on the surface of bulk substrates, such as $SrTiO_3$ (STO)[17] and (1-$x$)[$Pb(Mg_{1/3}Nb_{2/3})O_3$]-$x$[$PbTiO_3$] (PMN-PT)[18], and that of thin films such as $(Br,Sr)TiO_3$[19], $Pb(Zr,Ti)O_3$[10,16,20], STO[21], and $PbTiO_3$/STO superlattices.[22] Even though many interesting phenomena, including complicated hysteresis in current-voltage behaviors, have been reported in those devices, gate-voltage ($V_G$) scaling due to the influence of the high-$k$ dielectric thin film has not yet been explicitly demonstrated.

STO is a quantum paraelectric which exhibits a high, nonlinear dielectric constant.[26] It exhibits a low-lying-energy, soft-mode phonon, which results in a dielectric constant that is largely tunable with temperature, that is, from 200-300 at room temperature to a couple of thousands at low temperature.[27] Ideally, the high dielectric constant of STO can effectively screen the charges in nearby layers and also add tunability to STO-based heterostructures. Graphene-STO devices have hence been fabricated to understand the modifications of charge transport in graphene under a high-$k$ environment.[17,21] However, the quantum Hall effect, which is a hallmark of graphene, has not yet been observed on graphene-STO thin film heterostructures, possibly due to defect-induced leakage current through the STO thin films.. While quantum Hall effect was reported on a graphene-bulk STO substrate (0.5 mm thick) system, $V_G$ on the order of 10 V was required for the conventional device operation.[17] Therefore, a reasonably thin STO film is expected to work ideally as a gate dielectric for reduced operating voltage. Furthermore, the atomically flat surface of an epitaxial thin film helps to manifest graphene's intrinsic quantum transport properties that can serve as an indicator of the graphene quality on functional oxides.

In this Letter, we report voltage scaling of a graphene device on a high-$k$ dielectric, specifically, monolayer graphene on a 300 nm thick epitaxial STO thin film. The full electric analyses as a function of $V_G$, temperature, and magnetic field indicate that the reduction of operating voltage is successfully achieved without any degradation of the graphene device performance. Especially, the quantum Hall states of monolayer graphene on the atomically flat surface of an epitaxial STO thin film are demonstrated by the observation of a series of conductance quantization plateaus under magnetic fields, surviving even up to 200 K at 14 T. In addition, from considerations of the universality of quantum phenomena, the effective dielectric properties of an STO thin film in contact with graphene are deduced.

Epitaxial STO thin films were grown on atomically flat (001)-oriented Nb-doped (0.5 wt%) STO (Nb:STO) single-crystal substrates using pulsed laser epitaxy (PLE) at 700 ºC. A KrF excimer laser with a wavelength of 248 nm (IPEX 864, Lightmachinery, Nepean, Canada) was used to ablate STO ceramic target using a fluence of 1.3 J/cm$^2$ and a repetition rate of 5 Hz. In order to obtain stoichiometric STO thin films with minimized leakage current, we used a high oxygen partial pressure of 100 mTorr for the growth and postannealed the thin films at 400 ºC in 400 Torr of oxygen for an hour after the growth. Graphene flakes were mechanically exfoliated on a poly(methyl methacrylate) (PMMA)-coated silicon substrate. A monolayer graphene flake was identified by Raman spectroscopy and transferred onto the STO/Nb:STO substrate using the dry-transfer technique.[28] The source and drain electrodes consisting of a Cr/Au (5/50 nm) layer were patterned with conventional electron-beam lithography. The channel length and the width of the fabricated device were approximately 4 and 5.5 $\mu$m, respectively. For the transport experiment, the sample temperature and the external magnetic field were controlled by the cryostat (PPMS-Dynacool, Quantum Design Inc.), and the electrical measurements were carried out using a semiconductor parameter analyzer (B1500A, Agilent Technologies).

Figure 1a is a schematic diagram of our graphene-on-thin-film-STO device. The graphene layer, the epitaxial STO thin film, and the metallic Nb:STO single crystal substrate served as a channel, a gate dielectric, and a bottom gate electrode, respectively. Figure 1b shows the structural properties of the STO thin film. X-ray diffraction $\theta$-$2\theta$ scan around the (002) Bragg reflection reveals only the substrate peak without any Pendellösung fringes, indicating that our STO film is stoichiometric. In addition, atomic force microscopy (AFM) images, shown in the insets of Figure 2b, show clean surfaces with step heights equivalent to one unit cell of STO, which is TiO$_2$ layer terminated.[29] This step height is preserved after the 300 nm of STO thin film growth. We note that homoepitaxial growth of the STO thin film on a lattice-matched Nb:STO substrate enabled the simple transistor device geometry with minimized defect concentration. We also believe that the high surface quality of the STO thin film enabled the observation of quantum transport in graphene in an environment selectively influenced by a high-$k$ dielectric. The optical image of the completed device and the Raman spectrum of the exfoliated graphene flake are presented in Figure 1c.

The conductance without magnetic field, as a function of $V_G$ and temperature is shown in Figure 2a in the form of color map. For more detailed analyses the drain-current data versus $V_G$ at selected temperatures are plotted in Figure 2b. Here, the measurement range was determined by the level of gate leakage current, which was kept below 5 nA. As temperature decreases, the position of the charge-neutrality-point (CNP), corresponding to the minimum conductance, shifts toward a negative $V_G$ until the temperature reaches approximately 30 K. Below this temperature, the CNP is almost constant as indicated by the dark area in Figure 2a. In the high carrier density regime, marked by two rectangular boxes at the left and right sides of Figure 2b, the channel conductance is almost independent of the temperature, as shown in Figures 2c and 2e. This observation implies that temperature-related effects, such as phonon scattering or thermally activated carrier generation, are not relevant for the high carrier density regime. On the other hand, at low carrier densities near the CNP, marked with the rectangular box at the center of Figure 2b, a nontrivial temperature dependence is observed, as shown in Figure 2d. This indicates that temperature-sensitive physical processes like long-range Coulomb scattering due to charged impurities are involved in the charge transport of graphene near the CNP, as briefly discussed in the Supporting Information.[25,30-33]

At low enough temperatures under magnetic fields, graphene with high-mobility carriers is expected to exhibit quantum Hall conductance in a two-terminal device configuration.[34,35] Figure 3a shows the channel conductance quantization in our graphene-on-thin-film-STO device as a function of $V_G$ under various magnetic fields at 2 K. Here, the channel conductance $G_{CH}$ is equal to the inverse of channel resistance $R_{CH}$, which means the real channel resistance as distinguished from the value measured across the two-terminal device. The channel resistance is defined in the Supporting Information and illustrated in Figure S1. The observed conductance plateaus can be fitted well to the typical quantum Hall values of monolayer graphene, $G_{CH} = \nu\ (e^2/h)$, with typical filling factors of $\nu$ = 2, 6, 10, and 14. This conductance quantization phenomenon at different magnetic fields indicates that the essential electric characteristics of graphene remain intact even on top of the STO thin film. To test the stability under thermal fluctuation, conductance quantization was measured under various temperatures, as presented in Figure 3b. These data show that the signature of channel conductance quantization remains even up to 200 K. All these data indicate that conventional charge conduction through graphene in the transistor configuration is clearly reproduced, including the quantum Hall phenomena, but at much narrower $V_G$s because of the high-$k$ property of the epitaxial STO thin film.[36] Figure 3c displays the quantized channel conductance as a function of temperature and $V_G$ at a fixed magnetic field of 14 T, to illustrate the temperature evolution of the quantum Hall states.

It is worthwhile to note that there is no hysteresis in the conductance during the $V_G$ sweep in our high quality sample (see Figure 2 and Figure 3), which manifests a stark contrast to the transport characteristics reported previously on graphene-oxide thin film devices.[21,37] It seems that the conductance hysteresis in previous reports might have originated from other extrinsic effect related to the quality of both the graphene and STO thin film. Indeed, previous studies have used either CVD graphene on STO thin film[21] or a graphene device transferred from a silicon substrate to the target STO thin film substrate,[37] where the quantum transport phenomenon was not observed. In comparison with a conventional graphene device fabricated on a silicon/silicon-oxide substrate, our device shows nearly the same behavior without any noticeable differences other than the significant reduction of operating $V_G$ due to the high-$k$ property of the epitaxial STO thin film as originally designed.

One of the most attractive features in the use of high-$k$ thin-film gate-dielectrics is the possibility of ultrahigh doping into two-dimensional structured materials.[21] Owing to the small thickness as well as high-$k$ characteristic, the capacitance can be enlarged, which enables a much larger number of charge carriers to be injected into the conduction channel in comparison with the case of conventional silicon oxide gate dielectrics. Unfortunately, however, such advantages are limited by the leakage current as the gate dielectric layer becomes thinner. Therefore, to achieve ultrahigh doping of two-dimensional materials, an analysis of the origin of leakage current is necessary. Figure 4a presents leakage current characteristics of our graphene-on-thin-film-STO device taken during the measurements in Figure 2, where $V_G$ was applied to the Nb:STO single crystal substrate. An abrupt increase of gate-leakage current occurred for the positive $V_G$, and its threshold voltage was almost independent of temperature. For the negative

$V_G$, however, the threshold voltage increased relatively slowly in magnitude when the temperature decreased. Note that the polarity dependent leakage current characteristics can be understood based on simple band diagram, as discussed in Supporting Information (Figure S2).

The temperature-dependent threshold voltages of gate-leakage current in both bias directions are summarized in Figure 4b, together with the CNP voltage ($V_{CNP}$) of the graphene device from Figure 2a. Since it is well known that unintentional doping, for example, the occasional *p*-type doping due to PMMA residue remaining on the graphene surface, might occur in the graphene device, there is a possibility that the CNP might be located far away from the low-leakage current region. Fortunately in this case, as shown in Figure 4b, the CNP is positioned within the $V_G$ range of low leakage current level in the measured temperature range. Interestingly, it was reported that the work function difference between graphene and STO results in electron doping of graphene exfoliated on STO substrates,[38] which is consistent with our results.

For an in-depth understanding of the role of the epitaxial STO thin film as a substrate for a graphene device, the mechanism in which the dielectric characteristics couple with the contacted graphene layer should be clarified. Even though STO is known to have a high dielectric constant, its value changes depending on the structure or thickness of the layer as well as the temperature and the type of electrodes used.[17,21,27,39] For our sample, the variation of the STO dielectric constant with temperature was examined directly at a frequency of 1 kHz using the device shown in the inset of Figure 1c. Data are presented in Figure 4b along with the temperature dependent $V_{CNP}$. Although the top electrode of our sample is the mixture of graphene, Cr/Au pads, and silver paste contacts, the overall shape of the temperature dependence is similar to what has been reported in the literature.[39] Interestingly, the temperature dependence of the $V_{CNP}$ curve roughly follows that of the inverse of the dielectric constant (red circles) in the temperature region above 50 K, as shown in Figure 4b. This indicates indirect evidence that the shift of $V_{CNP}$ is correlated with the change in the STO dielectric properties, partly because of screening due to extrinsic doping caused by polymer residues on the graphene flake or charged impurities such as local polar nanoregions from the substrate.[40]

We note that the limited capacitance of thin-film STO dielectrics has been reported to be related to a low-permittivity interfacial "dead layer" that prohibits the full exercise of the high-$k$ material.[41] In addition, the capacitance of STO measured with the graphene electrode was lower than that with a metal electrode,[21] indicating that it is not plausible to use the measured dielectric constant values of bulk single-crystal STO for quantitative analysis of graphene transport. We therefore took an alternative approach to obtain information on the effective dielectric constant of STO contacted with graphene, using the universal behavior of conductance quantization under magnetic fields.[34,35,42] The successive plateaus of quantized conductance corresponding to quantum Hall states enable us to estimate the carrier density in the graphene channel as a function of $V_G$, which then gives the value of the effective dielectric constant of the epitaxial STO thin film. (See the Supporting Information for the detailed procedure of parameter extraction.) The effective dielectric constant $\varepsilon_{eff}$ was calculated to be between 250 and 330 from the proper matching of the $\nu = 6$ resistance plateau to the filling factor as presented in the lower inset of Figure 5. This estimate provides a slightly lower value than that in the inset of Figure 4b, which might be ascribed to the effect of poorly defined top contacts on the STO. The carriers induced in graphene are fully taken into account in the estimated results in Figure 5, while the mixed metallic contacts (described in the Supporting Information) are used for the direct measurement of dielectric constant in Figure 4b.

To estimate the temperature-dependent carrier density required to fill the $\nu = 2$ quantum Hall state, the voltage range $\Delta V_2$ corresponding to the $\nu = 2$ state was extracted at different temperatures. Here, $\Delta V_2$ is defined using the first derivative of the conductance with respect to $V_G$, as in the upper inset of Figure 5. $\Delta V_2$ shows little variation as the temperature changes from 2 to 50 K and starts to decrease slightly above 50 K, as shown in Figure 5. This also indicates that the effective dielectric constant below 50 K does not change much, which is approximately consistent with the results of the temperature dependence of $V_{CNP}$ (Figure 2a) as well as that of the STO dielectric constant (Figure 4b).

In conclusion, we demonstrated voltage scaling of a graphene layer on top of a high-$k$ STO thin film epitaxially grown on Nb:STO single-crystal substrate. Electrical transport in the device was controlled by the STO//Nb:STO back-gate with an atomically flat surface, and the transport data successfully confirmed the scaling-down of the operating voltage. No hysteretic behavior was observed in the conductance of the device during the $V_G$ sweep, assuring the high quality of our device. The quantum Hall effect was clearly observed in the two-terminal device, and the effective dielectric properties of STO in contact with graphene were deduced from the universality of the quantum Hall phenomena. Our study can be expanded to the combination of an atomically-flat high-$k$ thin-film dielectric and an atomically-thin two-dimensional semiconducting material, which offers the possibility for significant reduction of subthreshold swing for electronic device application.

FIGURES

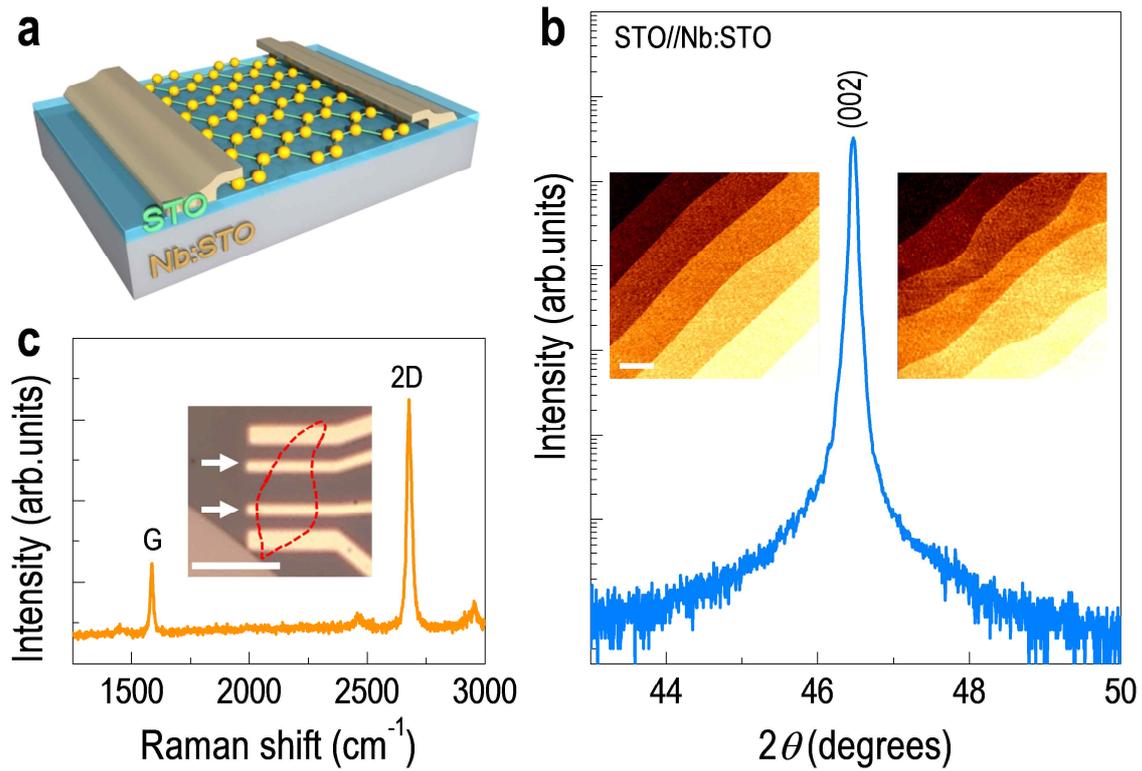

**Figure 1.** Device consisting of monolayer graphene on 300-nm-thick epitaxial STO thin film on Nb:STO substrate. (a) Schematic illustration of the device structure. (b) X-ray $\theta$-$2\theta$ scan of STO thin film on Nb:STO substrate. The insets show AFM images of the sample surface before (left) and after (right) STO film growth. The root-mean-square of surface roughness was typically around 0.6 nm. The scale bar corresponds to 500 nm. (c) Raman spectrum of the exfoliated graphene flake confirming the monolayer thickness. The inset is an optical image of the fabricated device. The red-dashed line indicates the shape of the exfoliated graphene flake. Two electrodes used for the electrical characterization are marked with white arrows. The scale bar corresponds to 10 $\mu$m.

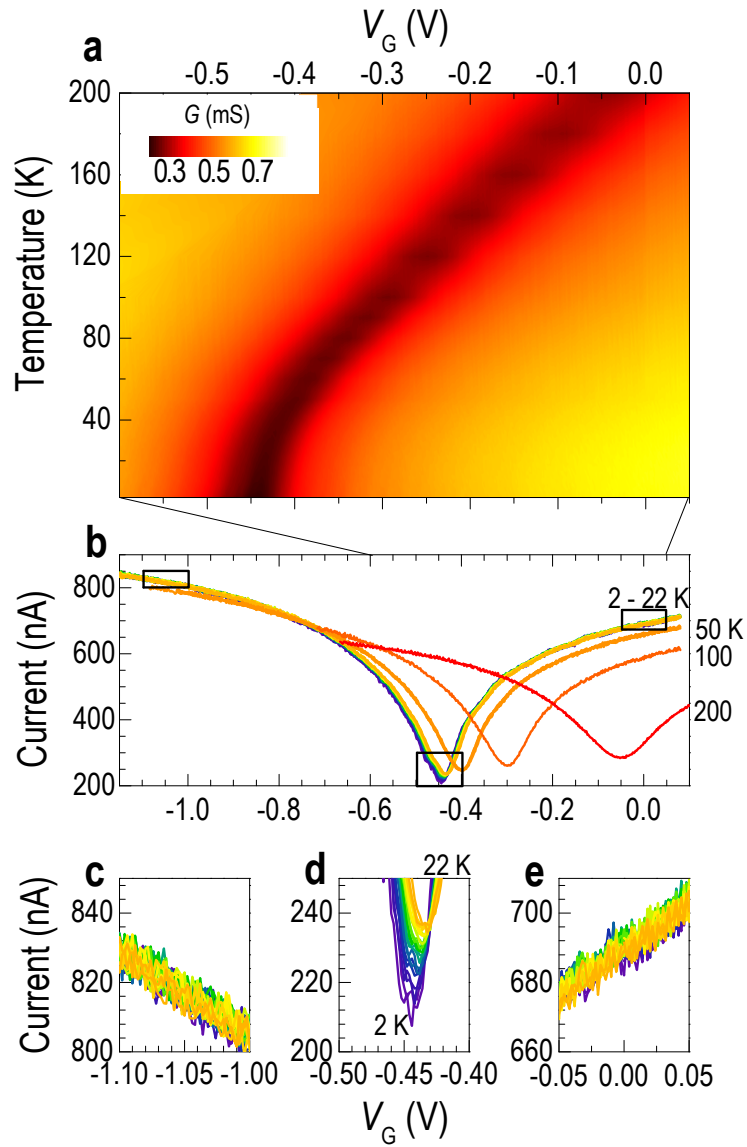

**Figure 2.** Temperature dependent conductance as a function of $V_G$. (a) 2D color map of conductance. The dark area corresponds to the low conductance state near the charge neutrality point. (b) Selected curves of $V_G$ versus current at $T = 200, 100, 50$ K, and selected values from 2 to 22 K. (c-e) Enlargements of the three different regions marked in (b) with red rectangular boxes, where the temperature varies from 2 to 22 K in 2 K intervals.

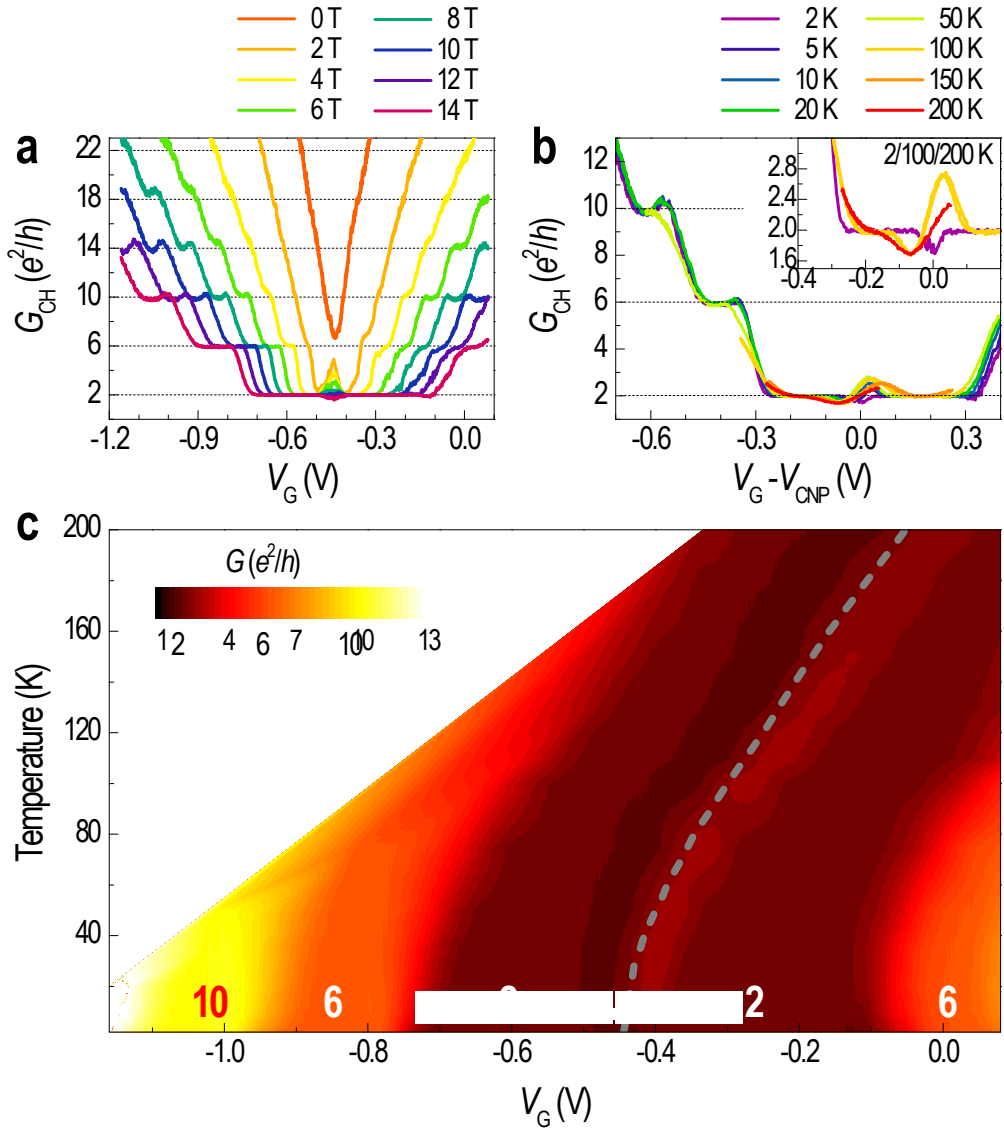

**Figure 3.** Channel conductance quantization as a function of $V_G$ (a) at $T = 2$ K under various magnetic fields and (b) under $H = 14$ T at various temperatures. See text for the definition of channel conductance. (c) 2D color map of channel conductance at a fixed magnetic field of $H = 14$ T. Gray-dashed line indicates the position of the charge neutrality point in the absence of magnetic fields.

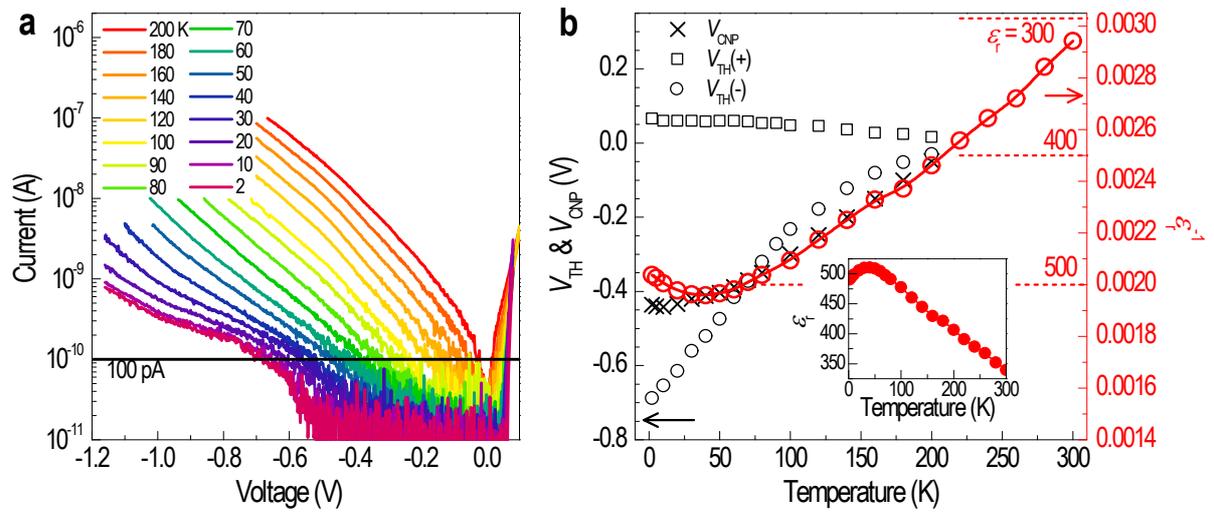

**Figure 4.** Leakage current and dielectric property of the graphene-thin-film-STO device. (a) Temperature dependent gate-leakage current as a function of $V_G$ during the measurement in Figure 2. (b) (Left axis) Temperature-dependent CNP and threshold voltages. The threshold voltage for the leakage current is defined as the gate voltage corresponding to a leakage current level of 100 pA. (Right axis) Temperature dependent inverse of the dielectric constant ($\varepsilon_r^{-1}$) of the STO thin film on Nb:STO substrate, directly measured from the device. The normal plot of $\varepsilon_r$ versus temperature is shown in the inset.

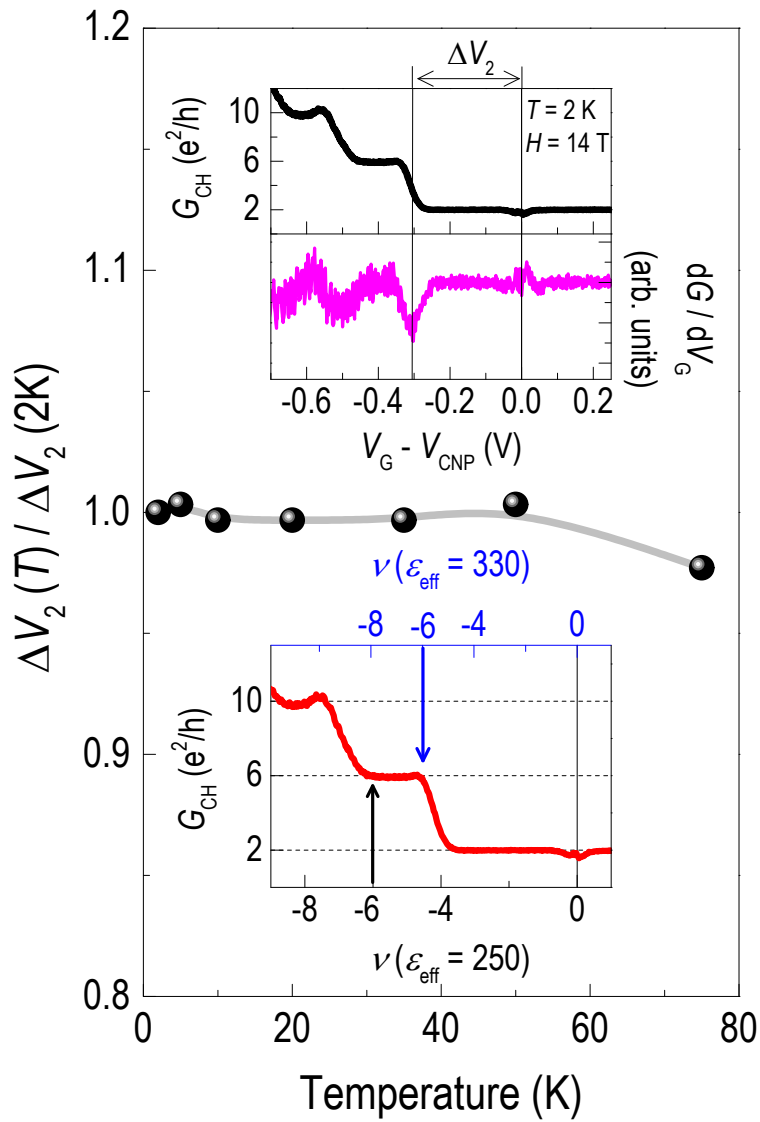

**Figure 5.** Estimation of effective dielectric constant and its temperature dependence. (Upper inset) Definition of $\Delta V_2$ corresponding to the $\nu = 2$ quantum Hall state, which does not change much at temperatures below $T = 50$ K. (Lower inset) Effective dielectric constant of STO thin film deduced from the quantum Hall plateau in the $\nu = 6$ state, which is in the range between $\varepsilon_{eff} = 250$ and $330$.

ASSOCIATED CONTENTS

**Supporting Information**

Estimation of contact resistance from two-terminal quantum Hall conductance; Estimation of the effective dielectric constant of STO from two-terminal quantum Hall conductance and its temperature dependence; Leakage current pathway of the graphene device on STO thin film.; Carrier-density-dependent conductance of graphene; Comparison of this work with other literatures. This material is available free of charge via the Internet at http://pubs.acs.org.

AUTHOR INFORMATION

**Corresponding Author**

*E-mail: choiws@skku.edu, energy.suh@skku.edu

**Author Contributions**

$^\perp$ These authors (J.P., H.K., K.T.K.) contributed equally to this work.

K.T.K. and W.S.C. prepared the high-quality epitaxial STO thin film. J.P., H.K., Y.Y., and D.S. fabricated the graphene device and did the electrical transport measurement. J.P., K.T.K, and W.S.C. analyzed the leakage current behavior. H.K., J.P., Y.H.L., and D.S. investigated the quantum Hall conductance. W.S.C., Y.H.L., and D.S. designed and led the project. All authors prepared the manuscript together.

**Notes**

The authors declare no competing financial interest.

ACKNOWLEDGEMENT

This work was supported by the Institute for Basic Science (IBS-R011-D1), Republic of Korea, and also supported by the Basic Science Research NRF-2015R1D1A1A01059850 (H.K.) and NRF-2014R1A2A2A01006478 (W.S.C.) through the National Research Foundation of Korea (NRF), funded by the Ministry of Science, ICT & Future Planning, Republic of Korea.

(36)  In Figure 3a, three quantum Hall plateaus corresponding to n = -2, -6, -10 under a magnetic field of 14 T are developed within the gate voltage range less than 1 V, in comparison with the device based on 300 nm $SiO_2$ where the gate voltage around 50 V is needed to see the quantum state for n = 10.

**TOC**

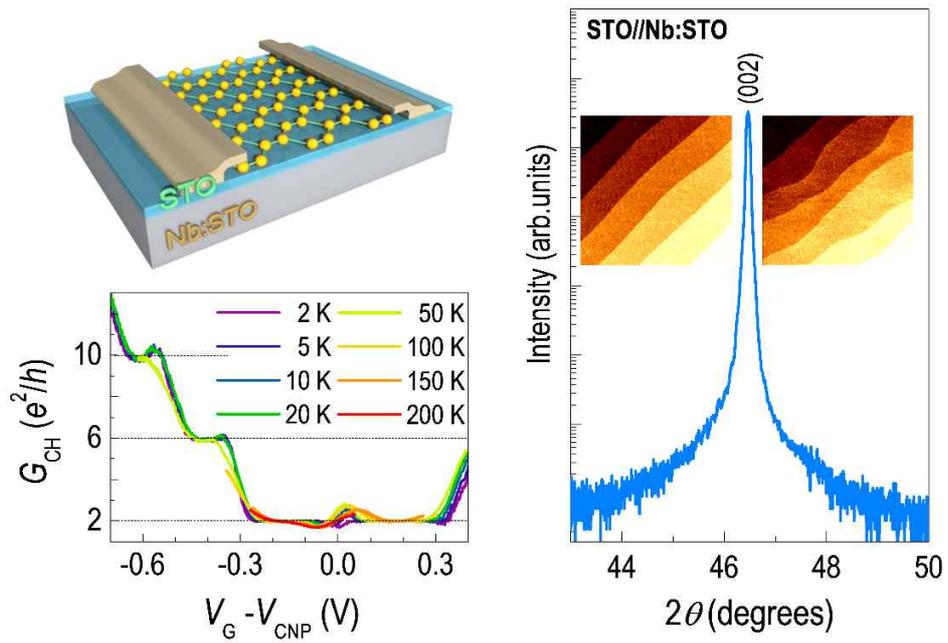

# Supporting Information

# Voltage Scaling of Graphene Device on SrTiO$_3$ Epitaxial Thin Film

*Jeongmin Park,[†,‡,⊥] Haeyong Kang,[‡,⊥] Kyeong Tae Kang,[†,§,⊥] Yoojoo Yun,[†,‡] Young Hee Lee,[†,‡,§] Woo Seok Choi,[§,\*] Dongseok Suh[†,‡,\*]*

[†] IBS Center for Integrated Nanostructure Physics, Institute for Basic Science, [‡] Department of Energy Science, [§] Department of Physics, Sungkyunkwan University, Suwon 440-746, Korea

[⊥] These authors contribute equally to this work.

[\*] Corresponding authors: choiws@skku.edu; energy.suh@skku.edu

**Estimation of contact resistance from two-terminal quantum Hall conductance**

For a deeper understanding of the effect of atomically flat, high-$k$ epitaxial thin film on the transport of graphene, we studied the quantum Hall conductance in this device. Even though the shape of the sample is not a Hall bar structure, the two-terminal quantum Hall conductance can be investigated at low temperatures under high magnetic fields. When the configuration of the two-terminal device is taken into account, several resistance components other than channel resistance ($R_{CH}$), such as wire resistance or contact resistance, could be added along the current pathway to give the measured resistance ($R_M$). Here, the sum of all subsidiary resistance components is regarded conceptually as a contact resistance ($R_C$), and it gives the following relation;

$$R_M = R_{CH} + R_C \ .$$

The measured data of conductance quantization under magnetic fields in our device is presented in Figure S1. We find that the resistance values of the quantized plateaus are different from the values corresponding to universal quantum Hall conductance that are expected in monolayer graphene. The difference is approximately 900 ohms for all quantized states, as indicated in Figure S1. If we assume that this difference originates from $R_C$, as suggested in the equation above, we can extract $R_{CH}$ by simply subtracting the $R_C$ value from $R_M$. Then the channel conductance becomes

$$G_{CH}(T,H) = \frac{1}{R_{CH}(T,H)} = \frac{1}{R_M(T,H) - R_C}$$

The $R_C$ value does not vary much from 2 K to 200 K.

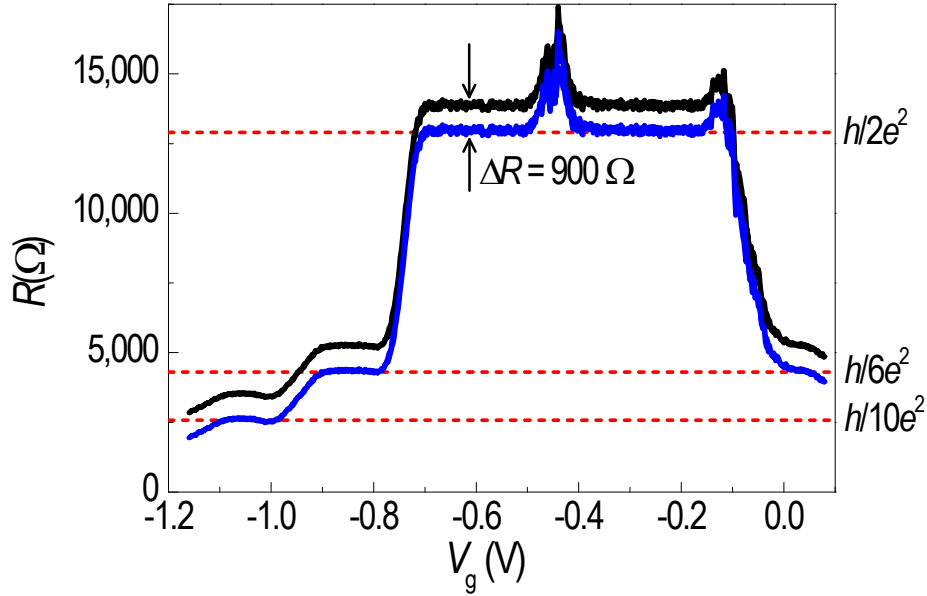

**Figure S1.** Comparison of experimental data in two-terminal resistance measurements under a magnetic field of 14 T with the theoretically expected quantized resistance values corresponding to quantum Hall states. The black curve indicates the measured data, and the blue curve is obtained after subtraction of 900 ohms from the measured data.

**Estimation of the effective dielectric constant of STO from two-terminal quantum Hall conductance and its temperature dependence**

The lower inset graph in Figure 5 indicates the method of using the quantum Hall plateaus to estimate the effective dielectric constant of the STO thin film. If we consider only the $\nu = 6$ state, the filling factor of this state should be matched with the quantized resistance plateau of $G_{CH} = 6$ ($e^2/h$), which results in the range of $\varepsilon_{eff}$ being approximately from 250 to 330 as presented in the lower inset of Figure 5.

For the temperature dependence of $\varepsilon_{eff}$, the location of the quantum Hall state transition from $v = 2$ to $v = 6$ was defined as $\Delta V_2$ relative to the charge neutrality point of $V_{CNP}$. For quantitative comparison, $\Delta V_2$ is obtained from the first derivative of the conductance with respect to gate voltage under magnetic field between quantized resistance plateaus as illustrated in the upper inset of Figure 5. Because the carrier number to occupy the $v = 2$ state at fixed magnetic field is independent of temperature, little variation of $\Delta V_2$ indicates that the effective dielectric constant also does not change much with temperature.

**Leakage current pathway of the graphene device on STO thin film**

For the leakage current path, there are three possibilities owing to the electrical contact made on the thin-film STO as seen in the inset of Figure 1c. The first is the Nb:STO/thin-film-STO/graphene, the second is the Nb:STO/thin-film-STO/Cr/Au, and the third is Nb:STO/thin-film-STO/silver-paste. In our work, it is not clear which one contributes most dominantly to the leakage current. In any case, Schottky contact behavior is clearly observed in the data presented in Figure 4a, which can be analyzed in terms of the schematic band diagrams shown in Figures S2a and S2b. The current flows relatively easily from Nb:STO to the top contact in the positive bias case (Figure S2a). On the other hand, for the negative bias, the current flow experiences a tunneling barrier due to a Schottky contact formed at the interface of the top contact and STO. For better performance with respect to leakage current, adoption of suitable conducting substrates for epitaxial STO in addition to a well-defined top contact area should be examined as options to achieve ultra-high doping in atomically thin two-dimensional materials.

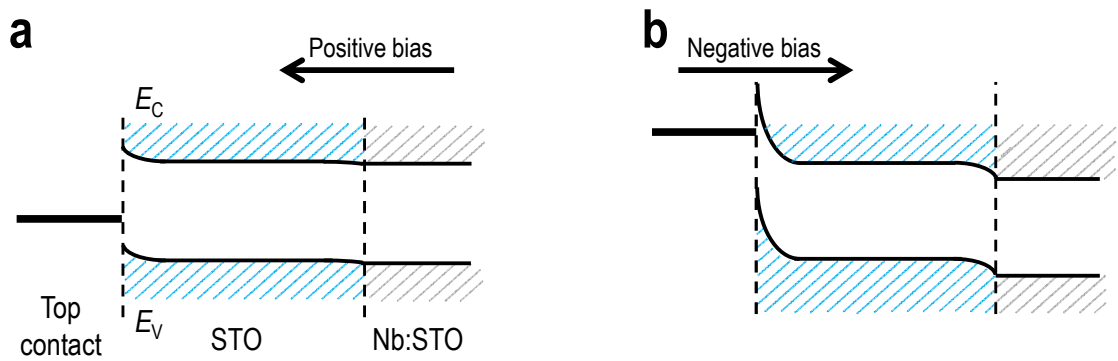

**Figure S2.** Schematic energy band diagram of the graphene-thin-film-STO device for (a) positive and (b) negative bias.

**Carrier density dependent conductance of graphene**

The conductance at 2 K shows a linear dependence in the vicinity of CNP but deviates from it as gate voltage increases as shown in Figure S3. Although there are still many debates on scattering mechanism affecting graphene transport, linear dependence on carrier density near CNP is considered to come from long-range Coulomb scattering.[25,30-33] The linear blue dotted lines in Figure S3 show that the conductance at low density is dominated by charged impurity scattering processes. As gate voltage increases, however, our data become fitted well with resonant scattering mechanism by strong and short impurity potential as indicated by red lines in Figure S3.[17,43-46] The conductivity in the resonant scattering dominant region follows

$$\sigma = \frac{2e^2}{\pi h} \frac{\alpha V_g}{n_i} \ln^2\left(\sqrt{\pi \alpha V_g}\, r\right),$$

where α is the ratio of carrier density induced by gate voltage ($n = \alpha V_g$), $n_i$ and $r$ are the concentration and the potential range of resonant defects, respectively. We can roughly estimate the density of resonant scatterers from the fitting parameters as $2 \times 10^{12}\, cm^{-2}$ where $r$ is set as 0.25 nm.

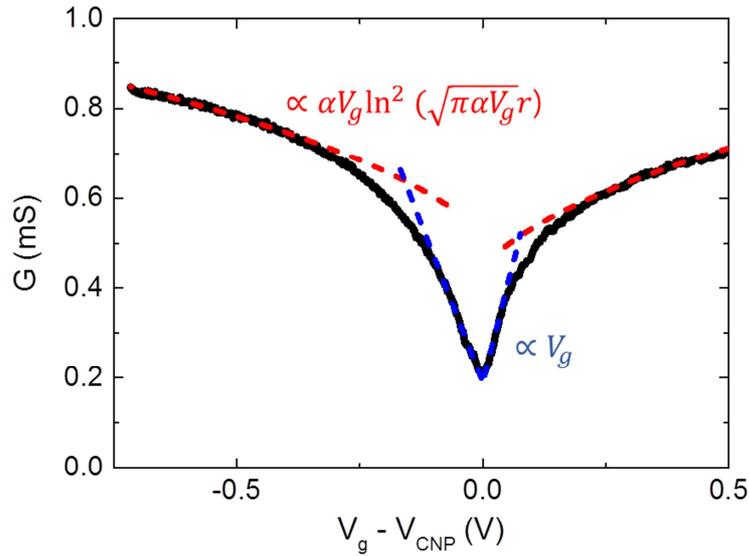

**Figure S3.** Gate voltage dependent conductance at 2 K. The blue and red dashed lines show linear and sublinear dependence on gate voltage, respectively.

**Comparison of this work with others**

The device structure and performance of this work and those of others in literature are summarized in the table below.

| Ref. | STO information | | | | | | Graphene transport | | | |
|---|---|---|---|---|---|---|---|---|---|---|
| | Growth method | $t$ [a] | Growth condition | | | Post-annealing | Hysteresis ($\Delta V_{CNP}$) [b] | Quantum Hall states | $\|V(2I_{CNP}) - V(I_{CNP})\|$ [c] | $\Delta V_2$ [d] |
| | | | $P(O_2)$ | $T$ | Laser fluence | | | | | |
| (Unit) | - | μm | mTorr | °C | J/cm² | - | V | - | V | V |
| This work | PLD | 0.3 | 100 | 700 | 1.3 | 400 Torr 400 °C, 1h | No hysteresis | Well-developed | 0.06 (2 K) 0.4 (200 K) | 0.3 (2 K) |
| [17] | bulk | 500 | - | - | - | - | No hysteresis | Well-developed | 0.6 (0.25 K) 3.7 (50 K) | 1.5 (0.25 K) |
| [21] | PLD | 0.3 | 100 | 700 | N/A | 500 Torr 400 °C, 1h | 1.5 (300K) | N/A | 0.8 (300 K) | N/A |
| [37] | PLD | 0.25 | N/A | N/A | N/A | 1100 °C, 6h | 1 (300K) 1.8 (4.2K) | N/A | 0.2 (4.2 K) 0.7 (300 K) | N/A |

(a) Thickness of STO.
(b) The difference between two CNP points depending on the direction of gate voltage sweep.
(c) The gate voltage required to increase the current at CNP twice.
(d) The gate voltage for quantum Hall state corresponding to ν=2.